\documentclass[12pt,letter]{article}

\usepackage{graphicx, epsfig, color}
\usepackage{amsmath,amssymb,hyperref}
\usepackage{subfigure}

\def\eslt{\not\!\!{E_T}}
\def\to{\rightarrow}

\def\bi{\begin{itemize}}
\def\ei{\end{itemize}}

\def\sps1ap{SPS1a$^\prime$}
\def\c1p{C1$^\prime$}

\def\tst{\tilde t}

\def\tg{\tilde g}

\def\tz{\widetilde Z}
\def\tw{\widetilde W}
\def\alt{\lesssim}
\def\agt{\gtrsim}
\def\be{\begin{equation}}  
\def\ee{\end{equation}}  
\def\bea{\begin{eqnarray}}  
\def\eea{\end{eqnarray}}  
\def\beas{\begin{eqnarray*}}  
\def\eeas{\end{eqnarray*}}

\newcommand{\hepph}[1]{hep-ph/#1}

\begin{document}
\begin{titlepage}
\begin{flushright}
OUHEP-180730,\ UH-511-1296-18
\end{flushright}

\vspace{0.5cm}
\begin{center}
{\Large \bf LHC luminosity and energy upgrades \\
confront natural supersymmetry models
}\\ 
\vspace{1.2cm} \renewcommand{\thefootnote}{\fnsymbol{footnote}}
{\large Howard Baer$^1$\footnote[1]{Email: baer@nhn.ou.edu }, 
Vernon Barger$^2$\footnote[2]{Email: barger@pheno.wisc.edu },
James S. Gainer$^3$\footnote[3]{Email: jgainer@hawaii.edu },\\
Dibyashree Sengupta$^1$\footnote[4]{Email: Dibyashree.Sengupta-1@ou.edu },
Hasan Serce$^1$\footnote[5]{Email: serce@ou.edu} and
Xerxes Tata$^{3,4}$\footnote[6]{Email: tata@phys.hawaii.edu }
}\\ 
\vspace{1.2cm} \renewcommand{\thefootnote}{\arabic{footnote}}
{\it 
$^1$Department of Physics and Astronomy,
University of Oklahoma, Norman, OK 73019, USA \\
}
{\it 
$^2$Department of Physics,
University of Wisconsin, Madison, WI 53706, USA \\
}
{\it 
$^3$Department of Physics and Astronomy,
University of Hawaii, Honolulu, HI 96822, USA \\
}
{\it 
$^4$Centre for High Energy Physics, Indian Institute of Science,
Bangalore, 560012, India \\
}
\end{center}

\vspace{0.5cm}
\begin{abstract}
\noindent 

The electroweak fine-tuning measure $\Delta_{\rm EW}$ allows for 
correlated SUSY soft terms as are expected in any ultra-violet complete
theory. Requiring no less than 3\% electroweak fine-tuning implies
upper bounds of about 360~GeV on all higgsinos, while top squarks 
are lighter than $\sim 3$~TeV and gluinos are bounded by $\sim 6-9$~TeV.
We examine the reach for SUSY of the planned high luminosity 
(HL: 3 ab$^{-1}$ at 14~TeV) and the proposed high energy 
(HE: 15~ab$^{-1}$ at 27~TeV)
upgrades of the LHC
%
%A clarification of the notion of electroweak naturalness in
%supersymmetric models has led to revised upper bounds on superparticle
%masses: while higgsino-like charginos and neutralinos may not stray too
%far from the weak scale and are bounded by about 300 GeV, highly mixed
%top squarks may range up to $\sim 3$ TeV while gluinos can range up to 6
%TeV or more at little cost to naturalness.  It is important to compare
%these target upper bounds from natural SUSY models against the reach of
%planned high luminosity (HL) and high energy (HE) LHC upgrades so that
%weak scale supersymmetry may either be discovered or falsified.  We
%examine the 
via four LHC collider search channels relevant for natural SUSY:
1. gluino pair production followed by gluino decay to third generation
(s)quarks, 2. top-squark pair production followed by decay to third
generation quarks and light higgsinos, 3. neutral higgsino pair
production with QCD jet radiation (resulting in monojet events with soft 
dileptons), and 4. wino pair production followed by
decay to light higgsinos leading to same-sign diboson production.  
%We
%update previous HE-LHC reach estimates to target energy $\sqrt{s}=27$
%TeV and integrated luminosity (IL) of 15 ab$^{-1}$.  We plot the range
%of parameter space points from 
We confront our reach results with upper limits on superpartner masses in
four natural SUSY models:
natural gravity-mediation via the 1. two- and 2. three-extra-parameter 
non-universal Higgs models, %(NUHM2 and NUHM3), 
3. natural mini-landscape models with generalized mirage mediation
% (nGMM) 
and 4. natural anomaly-mediation %(nAMSB). 
%We
%find that current LHC experiments have only just begun cutting into
%allowed natural SUSY parameter space for the first three channels.
We find that while the HL-LHC can probe considerable portions of natural
SUSY parameter space in all these models, the HE-LHC will decisively
cover the entire natural SUSY parameter space with better than 3\% fine-tuning.
%with no worse than a part in 30 electroweak fine-tuning.
\vspace*{0.8cm}
%\noindent PACS numbers: 12.60.Jv,14.80.Va,14.80.Ly

\end{abstract}

\end{titlepage}

\section{Introduction}
\label{sec:intro}

With the discovery of the Higgs boson in 2012\cite{lhc_h}, 
the CERN LHC has verified the particle content of the Standard Model (SM).
In spite of this impressive triumph, many physicists still expect new
physics to be revealed at LHC. 
The primary reason is the instability of the SM Higgs boson 
mass under radiative corrections if the SM is embedded into
a high scale theory (such as string theory).
Starting with the SM scalar potential
\be
V=-\mu^2\phi^\dagger\phi+\lambda (\phi^\dagger\phi )^2,
\label{eq:V}
\ee
one finds the Higgs mass, including leading radiative corrections 
cutoff at an energy scale $\Lambda$ (where new physics degrees of freedom not 
present in the SM become important), to be
\be
m_h^2\simeq 2\mu^2+\delta m_h^2, \nonumber
\ee
with\footnote{Quadratic divergences in the SM were studied by
  Veltman\cite{veltman}. While the use of a cutoff as a regulator is not 
gauge invariant, the coefficient of $\Lambda^2$ in Eq.~(\ref{eq:dmh2})
is independent of $\xi$ in $R_{\xi}$ gauges. For subtleties on the
regulation scheme dependence of the quadratic sensitivity of the Higgs
boson mass to high scale physics, see Ref.~\cite{ej}.}
\be
\delta m_h^2\simeq \frac{3}{4\pi^2}\left(-\lambda_t^2+\frac{g^2}{4}+
\frac{g^2}{8\cos^2\theta_W}+\lambda\right)\Lambda^2.
\label{eq:dmh2}
\ee Here, $\lambda_t$ is the top quark Yukawa coupling given in the SM by
$\lambda_t=\frac{gm_t}{\sqrt{2}M_W}$ , $g$ is the $SU(2)_L$ gauge
coupling, $\theta_W$ is the Weinberg angle and $\lambda$ is the Higgs
quartic coupling in the Higgs boson potential (\ref{eq:V}). 
The quadratic sensitivity of the SM  Higgs boson mass to new physics 
at the high scale $\Lambda$ embodies the fine-tuning problem of the SM.
 If the new
physics scale $\Lambda\gg 1$ TeV, then the {\it free} parameter $\mu^2$
will have to be accordingly fine-tuned to maintain the Higgs mass at its
measured value $m_h=125.09\pm 0.24$ GeV\cite{lhc_mhiggs}.  The
fine-tuning gets consequently more implausible as the theory cutoff
$\Lambda$ extends significantly  beyond the weak scale.  The need for
large fine-tuning suggests
%is regarded as symptomatic of some pathology or 
a missing ingredient in the underlying theory because otherwise
seemingly {\em independent} contributions to the Higgs boson mass would then
need to have large unexplained cancellations in order to yield its
measured value.

%result must since the
%predictivity of physical models necessitates that independent
%contributions to any observable quantity should be comparable to or less
%than its measured value.

Perhaps the most elegant and compelling resolution\cite{hier} of the
fine-tuning problem is to extend the underlying Poincar\'{e} spacetime
symmetries to the more general superPoincar\'{e} group. In the
supersymmetrized version of the SM, along with weak scale soft SUSY
breaking terms (the so-called Minimal Supersymmetric Standard Model or
MSSM\cite{wss}), the quadratic cutoff dependence seen in Eq.~(\ref{eq:dmh2}) is
absent, leaving only relatively mild but intertwined logarthmic
sensitivity to high scale physics.  
In addition to including a cure for the divergent Higgs mass, 
the MSSM receives impressive support from 
{\it data} via several different virtual effects:
\bi
\item the measured values of gauge coupling constants are consistent with
unification under renormalization group running within the MSSM\cite{drw,adf}
\item the measured value of the top quark mass is within the range
required to trigger a radiatively-driven breakdown of electroweak
symmetry\cite{rewsb}, and
\item the measured value of the Higgs mass fall squarely with the narrow
window of MSSM prediction\cite{mhiggs}, and in fact agrees with the
radiatively-corrected MSSM $m_h$ calculation provided top squarks
$(\tst_{1,2} )$ lie in the TeV range and are highly mixed by TeV-scale
trilinear soft terms\cite{h125}.  
\ei

The {\it natural} MSSM seemingly requires the existence of several
superpartners (those that have direct couplings to the Higgs sector)
with masses not too far beyond the weak scale as typified by
$m_{weak}\simeq m_{W,Z,h}\sim 100$ GeV\cite{wss}.  So far, searches by
LHC experiments have failed to find any superpartners leading to
simplified model gluino $(\tg )$ mass limits such as $m_{\tg}\agt 2$
TeV\cite{atlas_gl}\cite{cms_gl} and top-squark $(\tst_1)$ mass limits
such as $m_{\tst_1}\agt 1.1$ TeV\cite{atlas_t1,cms_t1} -- along with
considerably weaker limits on electroweakly interacting
superpartners. The widening mass gap between the weak scale and the soft
breaking scale has seemingly sharpened the issue of a {\it Little
Hierarchy}: how can it be that $m_{weak}\ll m_{soft}$ when the soft
breaking scale is supposed to determine the weak scale?  Naively, one
might expect $m_{weak}\sim m_{soft}$ absent again any fine-tuning.
Indeed, early estimates of naturalness or lack of fine-tuning within
SUSY models seemed to require $m_{\tg}\alt 350$ GeV and $m_{\tst_1}\alt
400$ GeV for no worse than $3\%$ fine-tuning\cite{bg,ac,dg}.  Some more
recent naturalness calculations seemed to require {\it three} third
generation squarks with mass below about 500 GeV\cite{kn}.  The contrast
between these naturalness bounds and current LHC mass limits might
indicate a need to fine-tune within the MSSM to maintain $m_{weak}\sim
100$ GeV which in turn may signal some pathology or missing ingredient
this time within the SUSY paradigm.

An issue with these estimates is that they ignore the possibility that
model parameters-- usually taken to be independent in order to parametrize 
our ignorance of SUSY breaking-- 
should be correlated (inter-dependent) in ultra-violet complete theories.
Such correlations can lead to automatic cancellations between
terms involving large logarithms: thus, ignoring this possibility can
easily lead to large over-estimates of the UV sensitivity of the theory
\cite{comp3,mt,dew}.

To allow for the fact that the underlying model parameters are expected to
be correlated, we adopt the very conservative fine-tuning
measure, $\Delta_{\rm EW}$~\cite{ltr,rns}.  The quantity $\Delta_{\rm
EW}$ measures how well the {\it weak scale} MSSM Lagrangian parameters
match the measured value of the weak scale.  By minimizing the MSSM weak
scale scalar potential to determine the Higgs field vevs, one derives
the well-known expression relating the $Z$-boson mass to the SUSY
Lagrangian parameters:
\be \frac{m_Z^2}{2} = \frac{m_{H_d}^2 +
\Sigma_d^d -(m_{H_u}^2+\Sigma_u^u)\tan^2\beta}{\tan^2\beta -1} -\mu^2
\simeq  -m_{H_u}^2-\Sigma_u^u(\tst_{1,2})-\mu^2 .
\label{eq:mzs}
\ee 
Here, $\tan\beta =v_u/v_d$ is the ratio of Higgs field
vacuum-expectation-values and the $\Sigma_u^u$ and $\Sigma_d^d$ contain
an assortment of radiative corrections, the largest of which typically
arise from the top squarks. Expressions for the $\Sigma_u^u$ and
$\Sigma_d^d$ are given in the Appendix of Ref. \cite{rns}.  Thus,
$\Delta_{\rm EW}$ compares the maximal contribution on the
right-hand-side (RHS) of Eq. \ref{eq:mzs} to the value of $m_Z^2/2$.  If
the magnitudes of the terms on the RHS of Eq.~(\ref{eq:mzs}) are
individually comparable to $m_Z^2/2$, then no unnatural fine-tunings are
required to generate $m_Z=91.2$ GeV. 
%Thus, the measure $\Delta_{\rm EW}$
%applies to weak scale SUSY models like the pMSSM, but also to high scale
%models where the soft terms are correlated (as expected in any model
%with a well-specified SUSY breaking sector).
 We have shown that once
appropriate inter-parameter correlations are properly taken into
account\cite{comp3,mt,dew}, then the traditonal fine-tuning measure\cite{bg},
$\Delta_{\rm BG}\equiv max_i|\frac{\partial\log m_Z^2}{\partial\log p_i}|$
indeed reduces to $\Delta_{\rm EW}$.

A utilitarian feature of the naturalness calculation is that it leads to
upper bounds on sparticle masses which in turn provide targets for
present or future colliding beam experiments which seek to discover
superpartners or
falsify the weak scale SUSY hypothesis. But for which values of
$\Delta_{\rm EW}$ is SUSY natural?  The original calculations of
Barbieri-Giudice used $\Delta_{\rm BG}<10$, or no less than $\Delta_{\rm
BG}^{-1}=10\%$ fine-tuning.  
%In Ref. \cite{upper}, the top ten
%contributions to $\Delta_{\rm EW}$ were displayed for increasingly tuned
%models and it was shown that tuning-- or unnaturalness-- turns on for
%values of $\Delta_{\rm EW}\sim 20$, or 5\% finetuning. To be
We will, more conservatively, adopt a value $\Delta_{\rm EW}<30$ (3.3\%
electroweak fine-tuning) as an upper bound on natural SUSY models.\footnote{
The onset of fine-tuning for $\Delta_{\rm EW}>20-30$ is visually 
displayed in Ref. \cite{upper}.} 
That this is a qualitatively different criterion is
driven home by the fact that it is possible to have the same model with
both $\Delta_{\rm EW} < 30$ and $\Delta_{\rm BG}> 3000$ (if the latter is
naively evaluated with multiple uncorrelated soft terms\cite{mt}).

Natural models with low electroweak fine-tuning ($\Delta_{\rm
EW}\alt 30$) exhibit the following features:  
\bi
\item $|\mu |\sim 100-350$ GeV\cite{Chan:1997bi,bbh}
(the lighter the better) 
where $\mu \agt 100$ GeV is required to accommodate LEP2 limits 
from chargino pair production searches.\footnote{We assume that the
superpotential $\mu$-term makes the dominant contribution to 
the higgsino mass.}
\item $m_{H_u}^2$ is driven radiatively to small-- not large--
negative values at the weak scale 
({\it radiatively-driven naturalness})\cite{ltr,rns}.
\item The top squark contributions to the radiative corrections
$\Sigma_u^u(\tst_{1,2})$ are minimized for TeV-scale highly mixed top
squarks\cite{ltr}. This latter condition also lifts the Higgs mass to
$m_h\sim 125$ GeV. For $\Delta_{\rm EW}\alt 30$, the lighter top
squarks are bounded by $m_{\tst_1}\alt 3$ TeV\cite{rns,upper}.
\item The gluino mass, which feeds into the top-squark masses at one-loop 
and hence into the scalar potential at two-loop order,
is bounded by $m_{\tg}\alt 6-9$ TeV\cite{rns,upper} (depending on the details
of the model).
\ei

These new sparticle mass bounds derived from the $\Delta_{\rm EW}$
measure lie well beyond current LHC search limits and allow for the
possibility that SUSY is still natural and still awaiting discovery. The
question then is: how far along are LHC SUSY searches on their way to
discovering or falsifying supersymmetry?  And what sort of LHC upgrade
is needed to either discover or falsify natural SUSY?
%or else to drive the final nail in its proverbial coffin?  
Indeed, recently the European Strategy Study 
has begun %in order 
to assess what sort of accelerator (or other
experiments) are needed beyond high-luminosity LHC (HL-LHC). 
One option is %the possibility of doubling 
to double
the field strength of the dipole steering magnets to 16 Tesla.
This
%which 
would allow for an energy upgrade of LHC to $\sqrt{s}=27$ TeV with
an assumed 15 ab$^{-1}$ of integrated luminosity (HE-LHC).  The goal of
this paper is to re-examine the SUSY
theory/experiment confrontation 
with a view to informing these questions about future experiments
and to examine what collider options are
needed to completely probe the natural SUSY parameter space.  In doing
so, we will confront four different natural SUSY models with updated LHC
limits from four SUSY search channels which are deemed most important
for discovering/falsifying natural supersymmetry.

The four natural SUSY models we examine here include the following:
\bi
\item Natural gravity-mediation as exhibited in the two- and three-extra
parameter non-universal Higgs model (nNUHM2 and nNUHM3)\cite{nuhm2}.
The NUHM2 model has parameter space $m_0,\ m_{1/2},\ A_0,\ \tan\beta,\
\mu,\ m_A$ which allows for the required light higgsinos since the
superpotential $\mu$ parameter is now a freely adjustable input
parameter so that the necessary naturalness requirement that $\mu\alt
350$ GeV is easily obtained.  The nNUHM2,3 models assume gaugino mass
unification which under MSSM  RG evolution leads to weak scale gauginos in the
mass ratio $M_1:M_2:M_3\sim 1:2:7$ while naturalness requires $\mu<
M_1<M_2<M_3$ so that a higgsino-like WIMP is the lightest SUSY particle
(LSP).  The nNUHM3 model has the added feature that first/second
generation matter scalars need not be degenerate with third generation
scalars. This sort of feature emerges in top-down SUSY models
such as the natural mini-landscape\cite{mini,miniland}.
\item Natural (generalized) anomaly-mediation or nAMSB adopts the usual
AMSB masses but also allows for {\it non-universal} bulk Higgs masses
$m_{H_u}$ and $m_{H_d}$ as compared to bulk matter scalar masses
$m_0$\cite{namsb}.  It also includes some bulk trilinear $A_0$ soft term
contributions.  The parameter space is then $m_0,\ m_{3/2},\ A_0,\
\tan\beta,\ \mu,\ m_A$.  The non-universal and trilinear bulk terms
allow for $m_h\simeq 125$ GeV while allowing for naturalness in the
spectra.  For nAMSB, the electroweakinos are oriented such that $\mu<
M_2<M_1<M_3$ at the weak scale.  The LSP in nAMSB is a higgsino-like LSP
instead of wino-like as is typically assumed.  For greater generality,
one may include as well separate first/second versus third generation
bulk matter scalar masses $m_0(1,2)$ and $m_0(3)$.
\item Natural generalized mirage mediation (nGMM) models\cite{mirage}, 
in which one expects comparable anomaly- and modulus/gravity-mediated contributions
to soft breaking terms.  The nGMM parameter space\cite{ngmm} is
$\alpha,\ m_{3/2},\ c_m,\ c_{m3},\ a_3,\ \tan\beta,\ \mu,\ m_A$, where
$\alpha$ parametrizes the relative modulus-to-anomaly-mediation
contributions and the $c_m$, $c_{m3}$ and $a_3$ are continuous
generalizations of previous discrete parameters related to modular
weights. Since gaugino masses unify at some intermediate (mirage) scale
$\mu_{mir}=e^{-8\pi^2/\alpha }m_{\rm GUT}$, then the gaugino masses are
{\it compressed} compared to NUHM2(3) so one expects $\mu\ll M_1\alt
M_2\alt M_3$. As an example, we examine the mini-landscape picture
taking $m_{3/2}\simeq m_0(1,2) > 2 \times m_0(3)$\cite{miniland}.
\ei 
These four
models have been encoded in Isajet v7.88\cite{isajet} which we use
for spectra generation and the $\Delta_{\rm EW}$ calculation.
%For each of the four models, we scan over the relevant parameter space
%which is consistent with current LHC sparticle mass constraints, with $m_h=125\pm 2$ GeV 
%(adopting $\sim\pm 2$ GeV theory error in our Higgs mass calculation).
%
For each of the four models, we scan over the whole parameter space
(with $\tan\beta : 3-60$) and accept solutions which are consistent with
current LHC sparticle mass constraints, with $m_h=125\pm 2$ GeV
(adopting $\sim\pm 2$ GeV theory error in our Higgs mass calculation).
We also require that solutions have $\Delta_{\rm EW}<30$ in order to
satisfy naturalness-- which amounts to a reasonable SUSY model
prediction for the magnitude of the weak scale. For the nGMM parameter
space, we require $\alpha$ to be positive (real mirage unification) and
$\alpha<40$ so that anomaly mediation is not highly suppressed.

%We list in Table \ref{tab:scan} the scan limits used for each model.
%{\bf Hasan to fill in}
%
%\begin{table}
%\centering %%% 
%\begin{tabular}{cccc}
%\hline
%nNUHM2 & nNUHM3 & nAMSB & mini-landscape \\
%\hline
%$m_0:x-y$ TeV & $m_0(1,2): x-y$ TeV & $m_{3/2}:x-y$ TeV & $m_{3/2}:x-y$ TeV \\
%$m_{1/2}:x-y$ TeV & $m_{1/2}: x-y$ TeV & $m_{3/2}:x-y$ TeV & $m_{3/2}:x-y$ TeV \\
%\hline
%\end{tabular}
%\caption{Scan limits for various parameters in the natural SUSY models considered here.
%We also take $\tan\beta:3-60$, $\mu:100-350$ GeV and $m_A: 0.3-10$ TeV.
%}
%\label{tab:scan}
%\end{table}

The four most important search channels for natural SUSY at the LHC or
its upgrades are the following.  
\bi
\item Gluino pair production $pp\to\tg\tg X$ followed by either two-body
gluino decay to top squarks $\tg\to\tst_1^* t,\ \tst_1\bar{t}$ or, if
these are closed, then gluino three-body decays to mainly third
generation quarks\cite{woodside}: $\tg\to t\bar{t}\tz_i,\ b\bar{b}\tz_i$
or $t\bar{b}\tw_j^+ +c.c.$.
\item Top squark pair production $pp\to\tst_1\tst_1^* X$ followed by
$\tst_1\to t\tz_i$ or $b\tw_j^+$\cite{stop}.
\item Higgsino pair production via $pp\to\tz_i\tz_j j, \tw_1\tz_i j, \tw_1\tw_1
j$ channels is unlikely to be visible above SM $Zj$
background because the signal to background ratio is just
1-2\%\cite{mono}.  However, the $pp\to\tz_1\tz_2 j$ channel (with contributions
from $pp \to \tw_1\tz_2 j$) , where $\tz_2\to\ell\bar{\ell}\tz_1$ with a
soft OS dilepton pair and where the hard initial state radiated jet
supplies a trigger, offers a promising search channel for low mass
higgsinos with $m_{\tz_{1,2}}\sim 100-300$ GeV\cite{lljMET}.  Indeed,
the LHC collaborations have presented their first results for this
search\cite{cms_z1z2,atlas_z1z2}, and it is especially encouraging that
the ATLAS collaboration is able to access a $\tz_2-\tz_1$ mass gap as small as
2.5~GeV.
\item Wino pair production $pp\to\tw_2^\pm\tz_{3\ {\rm or} \ 4} X$ followed by
$\tw_2\to W\tz_{1,2}$ and $\tz_{3\ {\rm or} \ 4}\to W^\pm\tw_1^\mp$. Half the
time, this final state leads to a {\it same-sign diboson} (SSdB) final
state which, when followed by leptonic $W$ decays, leads to same-sign
dileptons $+MET$ with very little accompanying jet activity\cite{ssdb}
(as opposed to SS dileptons arising from gluino cascade decays).  The
SSdB signature has very low SM background rates arising mainly from
$t\bar{t}W$ production.  
\ei 

In Sec.~\ref{sec:helhc}, we present our updated reach projections for
revised HE-LHC specifications with $\sqrt{s}=27$ TeV and a projected 
integrated luminosity (IL) of 15~ab$^{-1}$. 
In Sec. \ref{sec:confront}, we 
examine the four natural SUSY models introduced earlier and present LHC
bounds in each of these search channels, and also obtain reach
projections for HL- and HE-LHC.  We find that while HL-LHC can probe a
portion of natural SUSY parameter space, it will take an energy upgrade
to the HE-LHC option for a definitive search for natural weak scale
SUSY.  In Sec. \ref{sec:conclude}, we present a summary and conclusions.

\section{Updated reach projections of HE-LHC for
gluinos and top-squarks}
\label{sec:helhc}

In this section, we update previous HE-LHC reach analyses for 
top-squark pair production\cite{Baer:2017pba} and gluino pair 
production\cite{lhc33,Baer:2017pba} in natural SUSY
which were performed assuming $\sqrt{s}=33$ TeV and IL$=0.3-3$ ab$^{-1}$
to the updated values assigned for the European Strategy report, namely
$\sqrt{s}=27$ TeV and IL$=15$ ab$^{-1}$.
Along these lines, our first step is to generate updated total 
production cross sections for our signal processes.
\begin{figure}[tbp]
\begin{center}
\includegraphics[height=0.39\textheight]{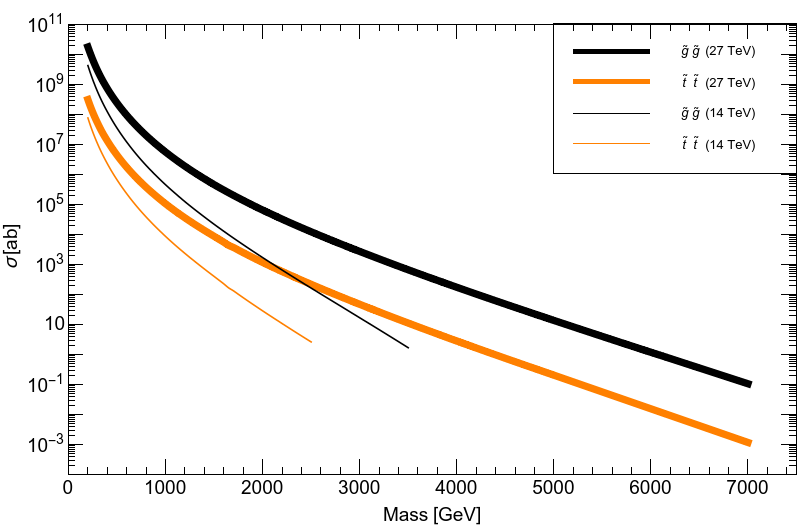}
\caption{Plot of NLL+NLO predictions\cite{Borschensky:2014cia} 
of $\sigma (pp\to \tg\tg X)$ and 
$\sigma (pp\to\tst_1\tst_1^* X)$ production at LHC for 
$\sqrt{s}=14$ and 27 TeV.
\label{fig:sigma}}
\end{center}
\end{figure}

In Fig. \ref{fig:sigma}, we plot the total production cross section for
$pp\to\tg\tg X$ (black) and $pp\to\tst_1\tst_1^* X$ (orange) at both 
$\sqrt{s}=14$ TeV (thin solid) and 27 TeV (thick solid). 
The results are computed at NLL+NLO and the 14 TeV results are taken from 
the study of Ref. \cite{Borschensky:2014cia} where we use the 
gluino pair production results for decoupled squarks.
Since Ref. \cite{Borschensky:2014cia} presents results for 
$\sqrt{s}=13$, 14, 33 and 100 TeV, we obtain total cross sections for
$\sqrt{s}=27$ TeV via interpolation of the 14 and 33 TeV results. 
Specifically, we fit $\log\sqrt{s}$ versus $\log\sigma_{tot}$ 
to a quadratic and used the resulting function to obtain 
$\sqrt{s}=27$ TeV cross sections.

From the results shown in Fig. \ref{fig:sigma}, we see that for 
$m_{\tg}=2$ TeV, then the gluino pair production cross section ratio 
$\sigma (27)/\sigma (14)=38$ while for $m_{\tg}=3.5$ TeV this ratio 
increases to $\sim 394$. For $m_{\tst}=1$ TeV, then we find a
total top squark pair production ratio $\sigma (27)/\sigma (14)=12$
while for $m_{\tst_1}=2.5$ TeV then $\sigma (27)/\sigma (14)$ increases
to 83. 
These ratios clearly reflect the advantage of moving to higher 
LHC energies in order to probe more massive strongly interacting sparticles.

\subsection{Updated top squark analysis for $\sqrt{s}=27$ TeV}
\label{sssec:stop27}

In Ref. \cite{Baer:2017pba}, the reach of a 33 TeV LHC upgrade for 
top-squark pair production was investigated. Here, we repeat the analysis
but for updated LHC energy upgrade $\sqrt{s}=27$ TeV. We use 
Madgraph\cite{madgraph} to generate top-squark pair production events within a
simplified model where $\tst_1\to b\tw_1^+$ at 50\%, and
$\tst_1\to t\tz_{1,2}$ each at 25\% branching fraction, which are typical
of most natural SUSY models\cite{stop}. The higgsino-like
electroweakino masses are $m_{\tz_{1,2},\tw_1^\pm}\simeq 150$ GeV.
We interface Madgraph with Pythia\cite{pythia} for initial/final state 
showering, hadronization and underlying event simulation. 
The Delphes toy detector  simulation\cite{delphes} is used with specifications
as listed in Ref. \cite{Baer:2017pba} (which we will not repeat here).
We also used Madgraph-Pythia-Delphes for a variety of SM background processes
which are listed in Table \ref{tab:stopBG}.
\begin{table}
\centering %%% 
\begin{tabular}{lc}
\hline
process & $\sigma$ (ab) \\
\hline
$b\bar{b}Z$ &       $1.87$ \\
$t\bar{t}Z$ &       $1.1$ \\
$t$         &       $4.4\times 10^{-2}$ \\
$t\bar{t}$  &       $3.3\times 10^{-2}$ \\
$t\bar{t}b\bar{b}$      &        $2.3\times 10^{-2}$ \\
$t\bar{t}t\bar{t}$      &        $1.7\times 10^{-3}$ \\
$t\bar{t}h$ &        $6.8\times 10^{-4}$ \\
total & $3.07$ \\
\hline
\end{tabular}
\caption{Cross sections in {\it ab} after cuts, 
listed in Sec.~\ref{sssec:stop27},
from SM background
processes  for the top-squark pair
production analysis at $\sqrt{s}=27$ TeV.  }
\label{tab:stopBG}
\end{table}

In Ref. \cite{Baer:2017pba}, an optimized set of cuts was found for
extracting the signal from a 2.75 TeV top squark over SM backgrounds
at $\sqrt{s}=33$ TeV LHC upgrade. The cuts that were settled upon were
\begin{itemize}
\item $n(b-jets)\ge 2$,
\item $n(isol.\ leptons)=0$,
\item $E_T^{miss}>max(1500\ {\rm GeV},0.2 M_{eff})$,
\item $E_T(j_1)>1000$ GeV,
\item $E_T(j_2)>600$ GeV,
\item $S_T>0.1$ and
\item $\Delta\phi (\vec{E}_T^{miss},{\rm jet\ close})>30$ deg.
\end{itemize}
In the above, $M_{eff}$ is the usual effective mass variable, $S_T$ is transverse
sphericity and the $\Delta\phi$ cut is on the transverse opening angle between
the missing $E_T$ vector and the closest jet (which helps reduce background from boosted tops in $t\bar{t}$ production).
The surviving background rates in ab are listed in Table \ref{tab:stopBG}. 
We use the same $K$-factors as listed in Ref. \cite{Baer:2017pba} to 
bring our total background cross sections into accord with various 
beyond-leading-order calculations. 
In the present analysis, we have also included  the $t\bar{t}Z$ background
calculation which was not present in Ref. \cite{Baer:2017pba}.

\begin{figure}[tbp]
\begin{center}
\includegraphics[height=0.39\textheight]{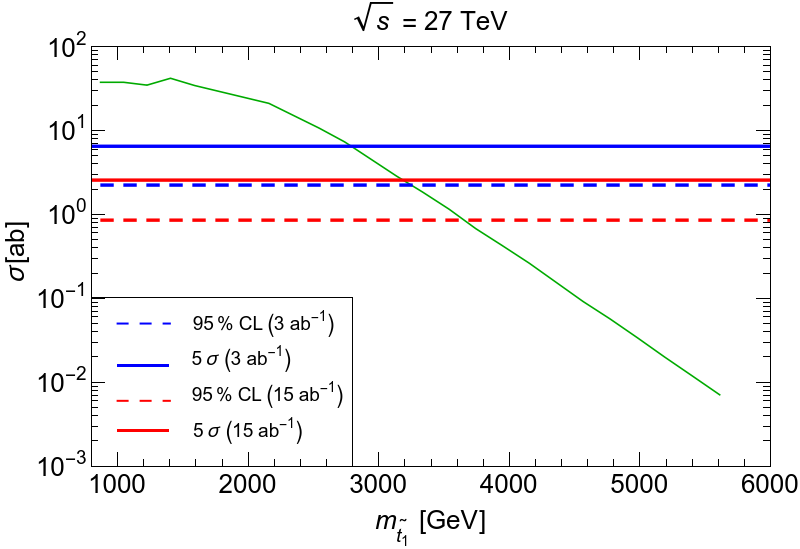}
\caption{Plot of top-squark pair production cross section vs. $m_{\tst_1}$ 
after cuts at HE-LHC with $\sqrt{s}=27$ TeV (green curve). 
We also show the $5\sigma$ and 95\% CL reach lines
assuming  3 and 15 ab$^{-1}$ of integrated luminosity (for a single detector).
\label{fig:stop}}
\end{center}
\end{figure}

Using these background rates for LHC at $\sqrt{s}=27$ TeV, we compute the
$5\sigma$ and $95\%$ CL reach of HE-LHC for 3 and 15 ab$^{-1}$ of
integrated luminosity using Poisson statistics. 
These results are plotted in Fig. \ref{fig:stop} along with the
top-squark pair production cross section after cuts versus $m_{\tst_1}$.
From the figure, we see the $5\sigma$ discovery reach of LHC27
extends to $m_{\tst_1}=2800$ GeV for 3 ab$^{-1}$ and to 3160 GeV for
15 ab$^{-1}$. The 95\% CL exclusion limits extend to $m_{\tst_1}=3250$ GeV
for 3 ab$^{-1}$ and to $m_{\tst_1}=3650$ GeV for 15 ab$^{-1}$. We see
that $S/B$ exceeds 0.8 whenever we deem the signal to be observable. 
Of course, somewhat increased reach limits can be obtained in the event 
of a combined ATLAS/CMS analysis. 

\subsection{Updated gluino analysis for $\sqrt{s}=27$ TeV}
\label{ssec:gluino27}

In Ref. \cite{Baer:2017pba}, optimized cuts were
investigated for extracting the signal from a 5.4 TeV gluino over 
SM backgrounds at a $\sqrt{s}=33$ TeV LHC upgrade. 
The optimized cuts were found to be
\begin{itemize}
\item $n(b-jets)\ge 2$,
\item $n(isol.\ leptons)=0$,
\item $E_T^{miss}>max(1900\ {\rm GeV},0.2 M_{eff})$,
\item $E_T(j_1)>1300$ GeV,
\item $E_T(j_2)>900$ GeV,
\item $E_T(j_3)>200$ GeV,
\item $E_T(j_4)>200$ GeV,
\item $S_T>0.1$ and
\item $\Delta\phi (\vec{E}_T^{miss},{\rm jet\ close})>10$ deg.
\end{itemize}
The corresponding backgrounds in ab after cuts are listed in 
Table~\ref{tab:gluinoBG}. The backgrounds are again normalized to recent
beyond-leading-order results as detailed in Ref. \cite{Baer:2017pba}.
We again compute the $5\sigma$ reach and $95\%$ CL exclusion lines using
Poisson statistics for 3 and 15 ab$^{-1}$ of integrated luminosity.
\begin{table}
\centering %%% 
\begin{tabular}{lc}
\hline
process & $\sigma$ (ab) \\
\hline
$b\bar{b}Z$ &       $0.061$ \\
$t\bar{t}Z$ &       $0.037$ \\
$t$         &       $0.003$ \\
$t\bar{t}$  &       $0.026$ \\
$t\bar{t}b\bar{b}$      &        $0.0046$ \\
$t\bar{t}t\bar{t}$      &        $0.0$ \\
$t\bar{t}h$ &        $8.1\times 10^{-4}$ \\
total       &        $0.132$ \\
\hline
\end{tabular}
\caption{Cross sections in {\it ab} after cuts, listed in
  Sec.~\ref{ssec:gluino27}, from SM background processes 
for the gluino pair production analysis at $\sqrt{s}=27$ TeV.
}
\label{tab:gluinoBG}
\end{table}

\begin{figure}[tbp]
\begin{center}
\includegraphics[height=0.39\textheight]{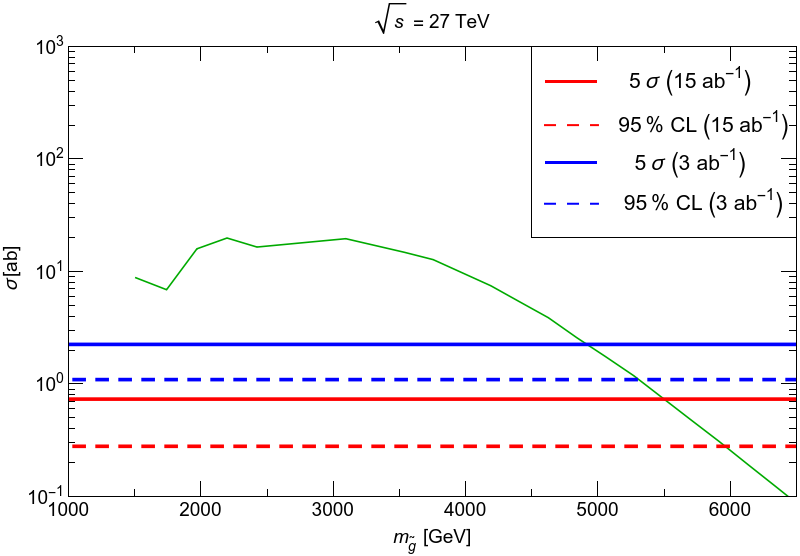}
\caption{Plot of gluino pair production cross section vs. $m_{\tg}$ 
after cuts at HE-LHC with $\sqrt{s}=27$ TeV (green curve). 
We also show the $5\sigma$ and 95\% CL reach lines
assuming 3 and 15 ab$^{-1}$ of integrated luminosity.
\label{fig:gluino}}
\end{center}
\end{figure}

Our results are shown in Fig. \ref{fig:gluino} where we plot the gluino pair
production signal versus $m_{\tg}$ for a nNUHM2 model line with 
parameter choice $m_0=5m_{1/2}$, $A_0=-1.6 m_0$, $m_A=m_{1/2}$,
$\tan\beta =10$ and $\mu =150$ GeV with varying $m_{1/2}$. We do not 
expect the results to be sensitive to this precise choice as long as first
generation squarks are heavy.
From the Figure, we see that the $5\sigma$ discovery reach of LHC27
extends to $m_{\tg}=4900$ GeV for 3 ab$^{-1}$ and to $m_{\tg}=5500$ GeV
for 15 ab$^{-1}$ of integrated luminosity. The corresponding $95\%$ CL exclusion
reaches extend to $m_{\tg}=5300$ GeV for 3 ab$^{-1}$ and to $m_{\tg}=5900$ GeV
for 15 ab$^{-1}$ of integrated luminosity.

\section{Confronting natural SUSY models at the LHC and its upgrades}
\label{sec:confront}

\subsection{Gluino pair production}
\label{ssec:gl}

\begin{figure}[tbp]
\begin{center}
\includegraphics[height=0.39\textheight]{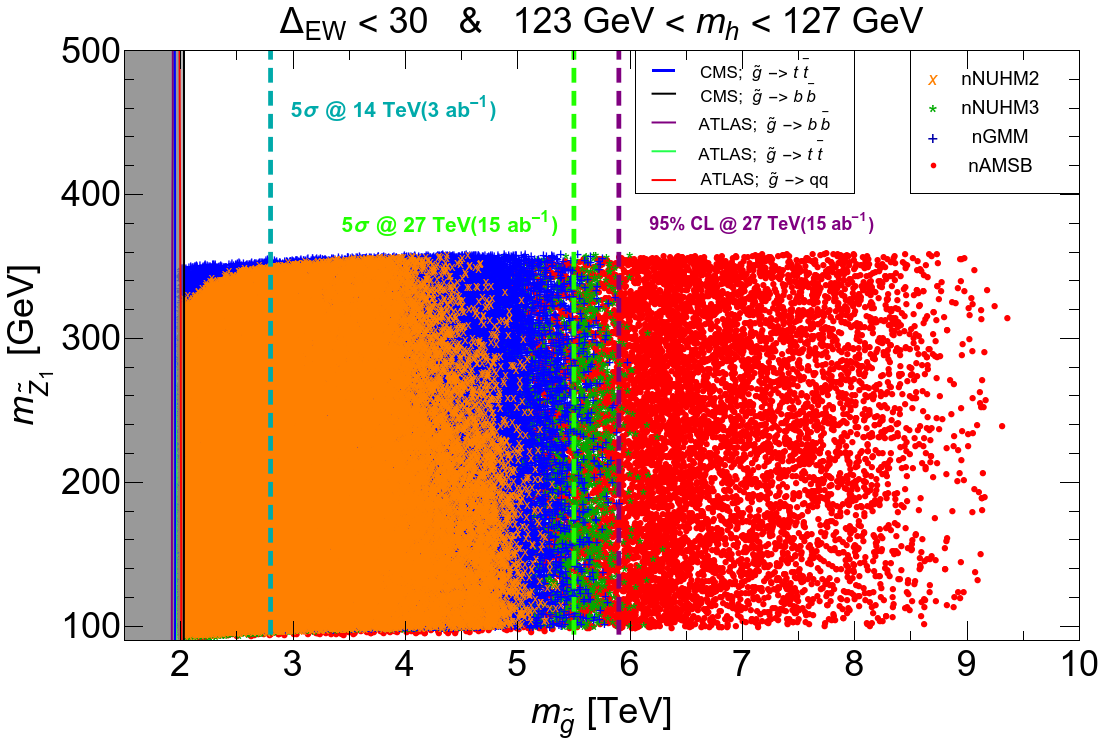}
\caption{Plot of points in the $m_{\tg}$ vs. $m_{\tz_1}$ plane 
from a scan over nNUHM2, nNUHM3, nGMM and nAMSB model parameter space.
We compare to recent search limits from the ATLAS/CMS experiments
(solid vertical lines) and future LHC upgrade options (dashed vertical lines).
\label{fig:mglmz1}}
\end{center}
\end{figure}

In Fig. \ref{fig:mglmz1} we display the results of our scans over
parameter space of the nNUMH2, nNUHM3, nAMSB and nGMM models with
$\Delta_{\rm EW}<30$ and with $m_h:123-127$ GeV in the $m_{\tg}$
vs. $m_{\tz_1}$ plane.  We also require $m_{\tg}>2$ TeV and
$m_{\tst_1}>1.1$ TeV in accord with recent simplified model mass limits
from ATLAS and CMS.  The density of points is not to be taken as
meaningful. Indeed, in a statistical study of IIB string theory
landscape\cite{dd}, it is argued that there should exist a power law
draw to large soft terms which would not be reflected here but which
would then favor larger sparticle masses beyond current LHC reach and
$m_h\simeq 125$ GeV.  The available natural parameter space can be
construed as some boundary enclosing all the natural SUSY scan points in
accord with the measured Higgs mass and current LHC sparticle mass
constraints.

From Fig. \ref{fig:mglmz1}, we see that the range of $m_{\tg}$ extends
from about 2 TeV to around $m_{\tg}\sim 6$ TeV for NUHM2,3 and nGMM
models but to significantly higher values for nAMSB. The upper limit on
$m_{\tg}$ occurs because the gluino mass drives top squark soft mass
terms to such large values that $\Sigma_u^u(\tst_{1,2})>30$, leading to
a violation of our naturalness criterion. To understand why higher
gluino masses are allowed in the nAMSB model, we first note that
$m_{\tg} \agt 6$~TeV occurs only for {\em negative} values of $A_0$. In
this case, in order to obtain $m_h$ consistent with its observed value
very large negative magnitudes of $A_0$ are required (compared to the
positive $A_0$ case).  The
resulting very large contribution of $A_t$ to their RG evolution then
strongly suppresses the weak scale soft top squark mass parameters,
allowing correspondingly larger values of $m_{\tg}$ ({\it vis \`a vis}
the other models). The fact that $|M_2|$ is smaller than $|M_3 |$ 
in the nAMSB case also helps.
%, but we suspect this is a sub-dominant effect because gluino
%masses beyond 6~TeV only occur for negative values of $A_0$. 
The range of $m_{\tz_1}$ varies from 100-350 GeV in accord with the range 
of $\mu$ which is bounded from below by LEP2 searches for chargino pair
production and bounded from above by naturalness in Eq.~(\ref{eq:mzs}).
We also show by the solid vertical lines around $m_{\tg}\sim 2$ TeV the
results of several ATLAS and CMS simplified model search limits for
gluino pair production\cite{atlas_gl}\cite{cms_gl}.  It is apparent from the plot that
a large range of parameter space remains to be explored.  The blue
dashed line around $m_{\tg}\sim 2800$ GeV shows the computed $5\sigma$
reach of high luminosity LHC (HL-LHC) with $\sqrt{s}=14$ TeV and 3
ab$^{-1}$ of integrated luminosity\cite{mgl}.  While the HL-LHC will
somewhat extend the SUSY search via the gluino pair production channel,
much of the allowed gluino mass range will remain beyond its reach.  We also
show with the green (purple) dashed lines the HE-LHC $5\sigma$ reach (95\%
CL exclusion region) for gluino pair production as computed in
Sec. \ref{sec:helhc} for $\sqrt{s}=27$ TeV and 15 ab$^{-1}$ of IL.  We
see that HE-LHC should probe nearly all of parameter space for the
nNUHM2, nNUHM3 and nGMM models while evidently a considerable fraction
of nAMSB parameter space would be beyond HE-LHC reach in the gluino pair
production channel.

\subsection{Top squark pair production}
\label{ssec:t1}

\begin{figure}[tbp]
\begin{center}
\includegraphics[height=0.39\textheight]{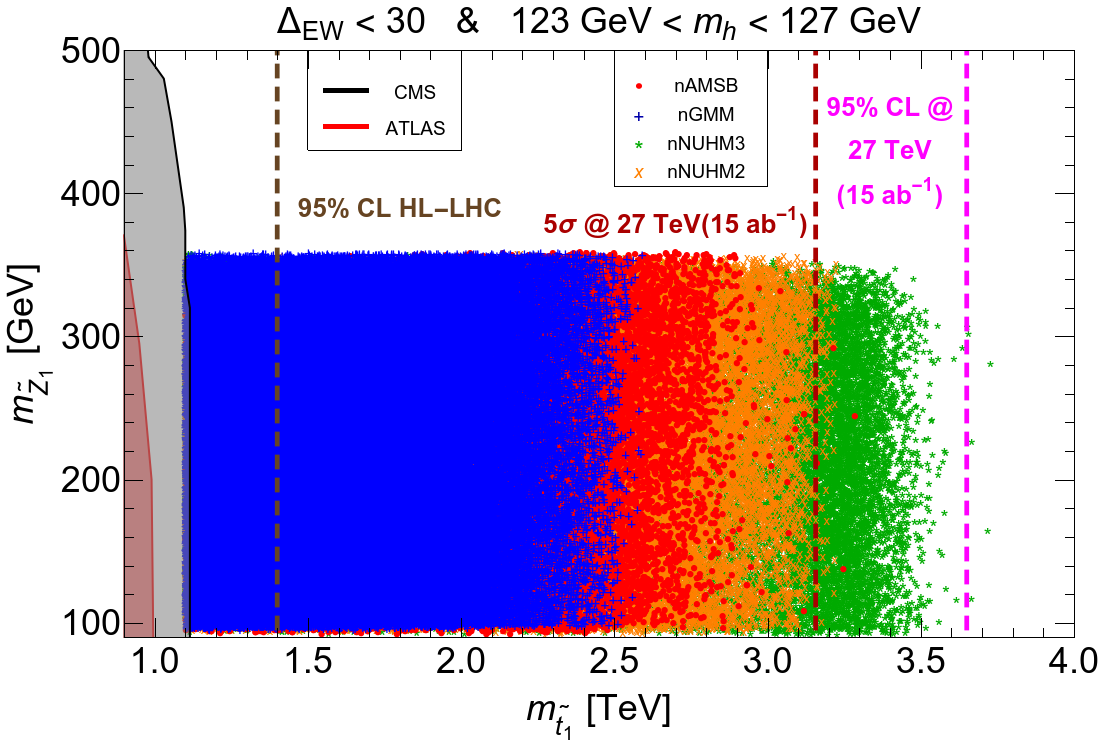}
\caption{Plot of points in the $m_{\tst_1}$ vs. $m_{\tz_1}$ plane 
from a scan over nNUHM2, nNUHM3, nGMM and nAMSB model parameter space.
We compare to recent search limits from the ATLAS/CMS experiments
(solid contours) and to projected future limits (dashed lines).
\label{fig:mt1mz1}}
\end{center}
\end{figure}

In Fig.~\ref{fig:mt1mz1}, we show the locus of scan points from the four
natural SUSY models in the $m_{\tst_1}$ vs. $m_{\tz_1}$ plane.  The
$m_{\tz_1}$ value is bounded by $\sim 350$ GeV so almost no points
occupy the near degeneracy region $m_{\tst_1}\sim m_{\tz_1}$ where much
LHC search effort has focussed.  We also show the current search limits
from ATLAS\cite{atlas_t1} and CMS\cite{cms_t1} as solid red and black
contours respectively.  These LHC search limits exclude some of natural
SUSY parameter space but evidently a large swath of natural SUSY
parameter space remains to be explored since top-squark masses may
extend up to $m_{\tst_1}\sim 3.5$ TeV without compromising naturalness.

The ATLAS collaboration projected 95\% CL exclusion region for top squarks at
HL-LHC\cite{atlas_t1_hl} is also shown by the black dashed line at
$m_{\tst_1}\sim 1.4$ TeV.  While HL-LHC will probe additional parameter
space, much of the top squark mass range will lie beyond its reach.  The
reach of HE-LHC with $\sqrt{s}=27$ TeV and IL of 15 ab$^{-1}$ was
computed in Sec. \ref{sec:helhc}.  We show the $5\sigma$ reach contour
as a red dashed line extending out to $m_{\tst_1}\sim 3.1$ TeV while the
95\% CL exclusion region extends to $m_{\tst_1}\sim 3650$ GeV.  The HE-LHC
apparently will be able to probe essentially the entire natural SUSY parameter space
in the top-squark pair production channel.

\begin{figure}[tbp]
\begin{center}
\includegraphics[height=0.39\textheight]{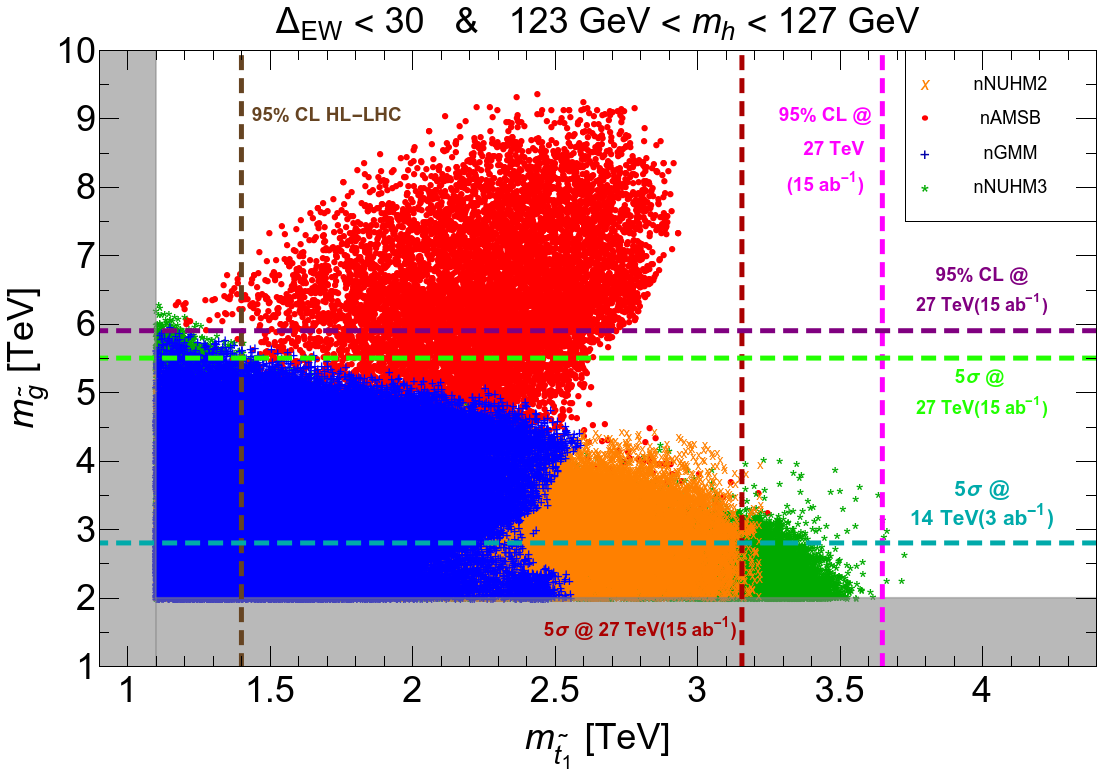}
\caption{Plot of points in the $m_{\tst_1}$ vs. $m_{\tg}$ plane 
from a scan over nNUHM2, nNUHM3, nGMM and nAMSB model parameter space.
We compare to projected future search limits from the LHC experiments.
\label{fig:mt1mgl}}
\end{center}
\end{figure}

In Fig. \ref{fig:mt1mgl} we show the gluino and top-squark reach values
in the $m_{\tst_1}$ vs. $m_{\tg}$ plane. The gray shaded region is excluded 
by the current search limits from CMS\cite{cms_gl}\cite{cms_t1}. 
In this plane, it is important
to note that in the nNUHM2, nNUHM3 and nGMM models, the highest values
of $m_{\tg}$ correspond to the lowest values of $m_{\tst_1}$ while the
highest $m_{\tst_1}$ values correspond to the lowest $m_{\tg}$ values.
Thus, a marginal signal in one of these channels (due to sparticle
masses being near their upper limit) should correspond to a robust
signal in the complementary channel. In particular, for nNUHM3 where
gluinos might be slightly beyond HE-LHC reach, the top squarks should be
readily detectable.  The nAMSB model case is different, because as we saw
in Sec.~\ref{ssec:gl}, the very large negative values of $A_0$ needed to
obtain the correct value of $m_h$ allow gluino masses in the $6-9$~TeV
range with modest values of $m_{\tst_1}$. (The top squark and gluino
mass values in the nAMSB model with $A_0> 0$ are in line with those in
the other models.)  We see that while gluino pair production might
escape detection at the HE-LHC in the nAMSB framework, the top squark
signal should be easily visible since $m_{\tst_1}\alt 3$ TeV in this
case.

\subsection{Higgsino pair production}
\label{ssec:hgsno}

The four higgsino-like neutralinos $\tw_1^\pm$ and $\tz_{1,2}$ are the
only SUSY particles required by naturalness to lie not too far above the
weak scale, $m_{weak}\sim 100$ GeV. In spite of their lightness, they
are very challenging to detect at LHC. The lightest neutralino evidently
comprises only a subdominant part of dark matter\cite{hgsno} and if
produced at LHC via $pp\to\tz_1\tz_1$ would escape detection.  In fact,
signals from electroweak higgsino pair production
$pp\to\tz_i\tz_j,\tw_1\tz_i,\tw_1\tw_1 + X$ ($i, j= 1,2$) are undetectable 
above SM backgrounds such as vector boson and top quark pair pruduction because
the decay products of the heavier higgsinos $\tw_1$ and $\tz_2$ are
expected to be soft.  The monojet signal arising from initial state QCD
radiation in higgsino pair production events has been evaluated in
Ref. \cite{mono} and was found to have similar shape distributions to
the dominant $pp\to Zj$ background but with background levels about 100
times larger than signal. See, however, Ref. \cite{hmw}.

%Meanwhile, production of the heavier higgsinos $\tw_1^\pm$ and $\tz_2$
%is also challenging. Charged higgsino pair production
%$pp\to\tw_1^+\tw_1^-$ followed by $\tw_1^-\to f\bar{f}^\prime \tz_1$
%gives rise to very soft particles in the final state since the bulk of
%$\tw_1^\pm$ energy goes into the $\tz_1$ rest mass. Thus, these final
%states will lie well below considerable backgrounds from processes like
%$WW$ and $t\bar{t}$ production.

\begin{figure}[tbp]
\begin{center}
\includegraphics[height=0.39\textheight]{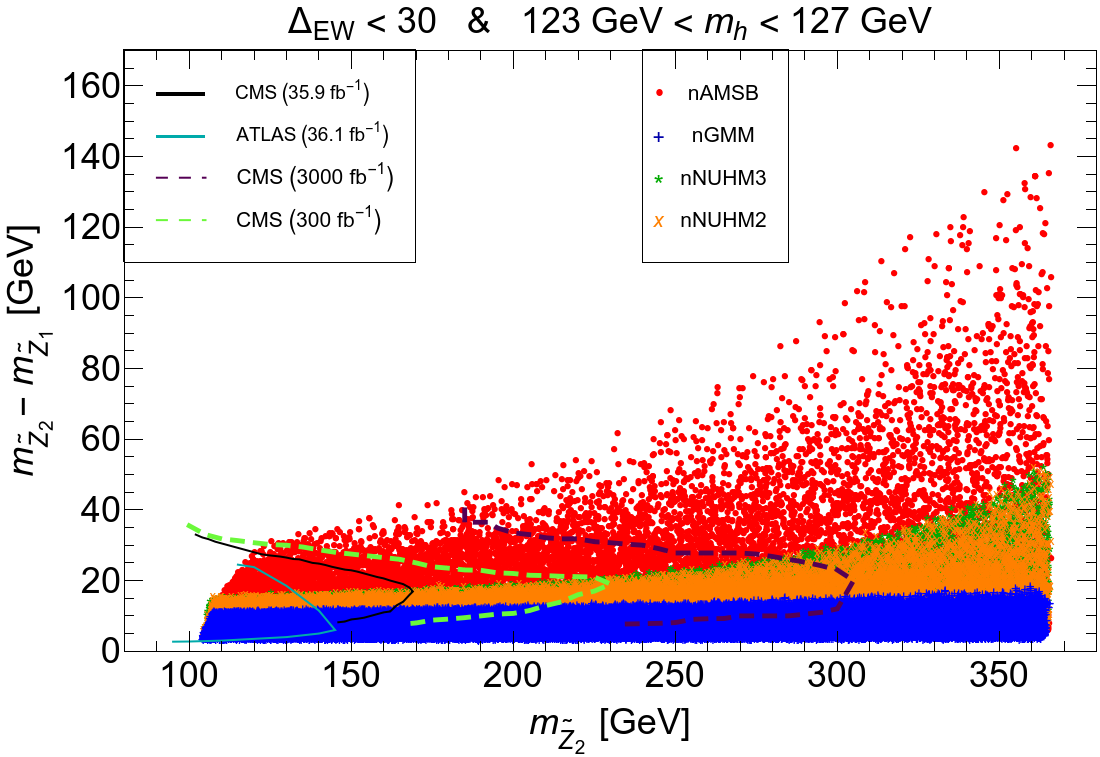}
\caption{Plot of points in the $m_{\tz_2}$ vs. $m_{\tz_2}-m_{\tz_1}$ plane 
from a scan over nNUHM2, nNUHM3, nGMM and nAMSB model parameter space.
We compare to recent search limits from the ATLAS/CMS experiments
and some projected luminosity upgrades as computed by CMS.
\label{fig:mz2mz1}}
\end{center}
\end{figure}

A way forward has been proposed via the $pp\to\tz_1\tz_2 j$ channel
where $\tz_2\to\ell^+\ell^-\tz_1$: a soft opposite-sign (OS) dilepton
pair recoils against a hard initial state jet radiation which serves as
a trigger\cite{lljMET}. Recent searches in this $\ell^+\ell^-j+\eslt$
channel have been performed by CMS\cite{cms_z1z2} and by
ATLAS\cite{atlas_z1z2}.  Their resultant reach contours are shown as
solid black and blue contours respectively in the $m_{\tz_2}$
vs. $m_{\tz_2}-m_{\tz_1}$ plane in Fig. \ref{fig:mz2mz1}.  These
searches have indeed begun to probe the most promising portion of the
parameter space, since the lighter range of $m_{\tz_2}$ masses have some
preference from naturalness.  The CMS experiment has also presented
projected exclusion contours for LHC14 with 300 fb$^{-1}$ and HL-LHC
with 3 ab$^{-1}$ shown as the green and purple dashed
contours\cite{cmstalk}.  We see that while these contours can probe
considerably more parameter space, much of natural SUSY parameter space
lies beyond these projected reaches. So far, reach contours for HE-LHC
in this search channel have not been computed but it may be anticipated
that HE-LHC will not be greatly beneficial here since $pp \to \tz_1\tz_2
j+X$ is primarily an electroweak production process so the signal cross
section will increase only marginally while QCD background processes
like $t\bar{t}$ production will increase substantially: harder cuts may,
however, be possible.  The nAMSB model inhabits typically a larger mass
gap region of the plane since in this model winos are much lighter than
in nNUHM2 or nGMM for a given gluino mass.  It is imperative that future
LHC searches try to squeeze their reach to the lowest
$m_{\tz_2}-m_{\tz_1}$ mass gaps which are favored to lie in the 3-5 GeV
region for string landscape projections\cite{dd}.
% in the nNUHM2 model. 
%

\subsection{Wino pair production}
\label{ssec:ssdb}

The wino pair production reaction $pp\to\tw_2^\pm\tz_4 X$ (in nNUHM2,3
and nGMM) or $pp\to\tw_2^\pm\tz_3 X$ (in nAMSB) offers a new and lucrative
search channel which is not present in unnatural models where $|\mu| \gg
M_{gauginos}$.  The decay modes $\tw_2^\pm\to W^\pm\tz_{1,2}$ and
$\tz_{3\ {\rm or}\ 4}\to W^\pm\tw_1^\mp$ lead to a same sign diboson
(SSdB) plus $\eslt$ final states accompanied by minimal jet activity- just
that arising from initial state radiation\cite{ssdb}. Thus, the ensuing
same-sign dilepton+$\eslt$ signature is quite different from that which
arises from gluino and squark pair production where multiple hard jets
are expected to be present. The SSdB signature from wino pair production
has very low SM backgrounds which might arise from processes like
$t\bar{t}W$ production.

\begin{figure}[tbp]
\begin{center}
\includegraphics[height=0.385\textheight]{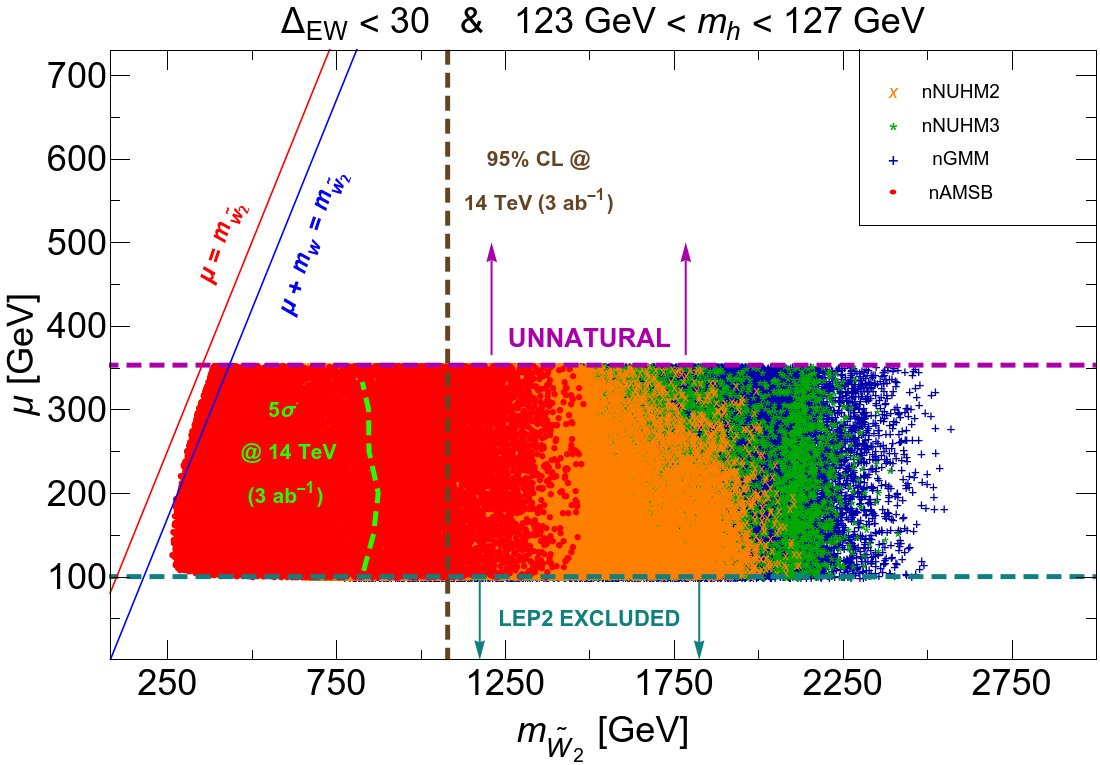}
\caption{Plot of points in the $m_{\tw_2^-}$ vs. $\mu$ plane 
from a scan over nNUHM2, nNUHM3, nGMM and nAMSB model parameter space.
We compare to projected search limits for the ATLAS/CMS experiments
at HL-LHC.
\label{fig:mw2mu}}
\end{center}
\end{figure}

In Fig. \ref{fig:mw2mu} we show the location of natural SUSY model
points in the $m_{\tw_2}$ vs. $\mu$ plane. The region with large $\mu$
is increasingly unnatural as indicated in the plot. From
Fig. \ref{fig:mw2mu}, we see that the nAMSB model points tend to
populate the lower $m_{\tw_2}$ region, $m_{\tw_2}\alt 1400$ GeV. This
is because $M_2\sim m_{\tg}/7$ in AMSB models with $m_{\tg}\alt 6-9$ TeV
from  naturalness considerations.
%
%have some degree of separation compared 
%to nNUHM2 and nNUHM3 in that they are
%bounded from above by 
%

We are unaware of any LHC search limits via the SSdB channel, though
this signature should begin to be competitive with the conventional $\eslt$
searches for an integrated luminosity of $\sim 100$~fb$^{-1}$ expected
to be accumulated by the end of LHC Run 2.  The projected HL-LHC
reach has been evaluated in Ref.~\cite{ssdb} where the $5\sigma$
discovery and $95\%$ CL exclusion dashed contours are shown.  Evidently
HL-LHC will be able to probe a large part of parameter space for the
nAMSB model while only a lesser portion of natural parameter space of
nNUHM2, nNUHM3 and nGMM models can be probed.  The corresponding reach
of HE-LHC has not been computed for the SSdB channel.  But again, since
this is an EW production channel, the signal rates are expected to rise
by a factor of a few by moving from $\sqrt{s}=14$ TeV to $\sqrt{s}=27$
TeV while some of the QCD backgrounds like $t\bar{t}$ production will
rise by much larger factors. We also note that because the heavy winos
are expected to decay to higgsinos plus a $W^\pm, Z$ or $h$ in the ratio
2:1:1\cite{ssdb}, $VV, Vh$ and $hh$ plus $\eslt$ signals may be
present, possibly with additional soft leptons from higgsino decays. A
study of these signals is beyond the scope of the present analysis.

\section{Summary and conclusions}
\label{sec:conclude}

Our goal, in this paper, was to ascertain what sort of LHC upgrades
might be sufficient to either discover or falsify natural supersymmetry.
We focused here on natural SUSY spectra 
%since these models are 
consistent with the measured value of the weak scale $m_{weak}\sim 100$
GeV without a need for implausible fine-tuning of model parameters.
Naturalness, after all, remains one of the major motivations for weak
supersymmetry and unnatural models seem highly implausible.
%The
%natural SUSY models-- with their simplicity, elegance and naturalness--
%stand a good chance of describing nature at the TeV scale and beyond.
To this end, we scanned over four different natural SUSY models: nNUHM2,
nNUHM3, nAMSB and nGMM. We obtained uppper limits on top squark masses
($m_{\tst_1}\alt 3.5$ TeV), gluino masses ($m_{\tg}\alt 6$ TeV in
nNUHM2,3 and nGMM, but $m_{\tg}\alt 9$ TeV in nAMSB) and higgsino and
wino masses.

We compared these against current LHC constraints and found large
regions of natural SUSY parameter space remain to be explored.  We also
compared against the HL-LHC upgrade: the HL-LHC with $\sqrt{s}=14$ TeV
and 3 ab$^{-1}$ of integrated luminosity will explore deeper into
natural SUSY parameter space but, barring a SUSY discovery, much of
parameter space will remain to be explored.  
We also updated the HE-LHC reach using the revised energy and integrated
luminosity targets as suggested by the ongoing European Strategy study:
$\sqrt{s}=27$ TeV and IL$=15$ ab$^{-1}$.  For these latter values, we
find a HE-LHC reach in $m_{\tst_1}$ to 3200 GeV at $5\sigma$ and $3650$
GeV at 95\% CL.  For the gluino, we find a HE-LHC reach to
$m_{\tg}=5500$ GeV at $5\sigma$ and 6000 GeV at 95\% CL. The gluino (top
squark) reach is reduced by about 600~GeV (400~GeV) if the integrated 
luminosity is instead  3~ab$^{-1}$. 

Comparing these values with upper limits from naturalness, we find the
HE-LHC is sufficient to probe the entire natural SUSY parameter space in the
top-squark pair production channel and also to almost
explore nNUHM2,3 and nGMM models in the gluino pair channel. Within
these models it is,
therefore, very likely that signals from top squark and gluino pair
production will be present at the HE-LHC. In the nAMSB model, it appears
that gluinos may be beyond the HE-LHC reach. 

% while the higher
% allowed gluino masses for nAMSB could be beyond HE-LHC reach.

We also compared the soft OS dilepton+jet signal from higgsino pair
production to current and future reach projections for HL-LHC. For this
channel, it will be important to explore neutralino mass gaps
$m_{\tz_2}-m_{\tz_1}$ down to $\sim 3$ GeV and higgsino masses up to
$\sim 350$ GeV for complete coverage.  We caution that the energy
upgrade of the LHC may not be as beneficial for this discovery channel
since QCD backgrounds are expected to rise more rapidly with energy than
the EW higgsino pair production signal channel. We also examined the
SSdB signature arising from charged and neutral wino pair production.
The HL-LHC may explore a portion of -- but not all of -- natural SUSY
parameter space in this channel. It is again unclear whether an energy
upgrade will help much in this channel since QCD backgrounds are
expected to increase more rapidly than the EW-produced signal channel
for an assumed wino mass $m_{\tw_2^\pm}$. We note, though, that there may be
signals from wino pair production in $VV$, $Vh$ and $hh$ + $\eslt$
channels which may also be interesting to explore.

To sum up: the key theoretical motivation for {\em weak scale} supersymmetry
as the stabilizer of the Higgs sector still remains, 
once we acknowledge that model parameters which are usually taken to be 
independent in spectra computer codes are expected to be correlated
in any ultraviolet complete theory. Our final assessment is that
the search for natural SUSY will, and should,  continue
on at LHC and HL-LHC, where more extensive regions of parameter space
may be explored.  The envisioned HE-LHC upgrade to $\sqrt{s}=27$ TeV and
IL$=15$ ab$^{-1}$ seems sufficient to either discover or falsify natural
SUSY in the top-squark pair production signal channel, very possibly
with an additional signal in the gluino-pair production channel.
It is possible that observable signals
may also emerge in the wino-pair or higgsino-pair plus monojet search
channels as well.
% although these latter channels are not guaranteed.

\section*{Acknowledgments}

This work was supported in part by the US Department of Energy, Office
of High Energy Physics. The computing for this project was performed at
the OU Supercomputing Center for Education \& Research (OSCER) at the
University of Oklahoma (OU). XT thanks the Centre for High Energy
Physics, Indian Institute of Science Bangalore, where part of this work
was done for their hospitality, and also the Infosys Foundation for
financial support that made his visit to Bangalore possible. 

%
%%%%%%%%%%%%%%%%%%%%%%%%%%%%%%%%%%%%%%%%%%%%%%%%%%%%%%

%
\end{document}